# Short Wave upwelling Radiative Flux (SWupRF) within NIR range for the selected greenhouse wavelength bands of $O_2$, $H_2O$, $CO_2$ and $CH_4$ by Argus 1000 along with GENSPECT line by line radiative transfer model.

**Rehan Siddiqui [1] , Rajinder Jagpal [1], Naif Al Salem[1], Brendan M. Quine [1, 2]**

[1]  Department of Physics and Astronomy, York University, 4700 Keele Street, Toronto, Canada, M3J 1P3: rehanrul@yorku.ca, jagpalraj@gmail.com, naif11@yorku.ca, bquine@yorku.ca.

[2]  Department of Earth and Space Science and Engineering, York University, 4700 Keele Street, Toronto, Canada, M3J 1P3: bquine@yorku.ca

## Abstract

This new study develops an algorithm for Short Wave upwelling Radiative Flux (SWupRF) for the spectral variations within near infrared (NIR) from 1100 to 1700 nm  wavelength band based on remote sensing data set of Argus 1000 micro-spectrometer observations. We calculate the SWupRF by investigating the total radiative flux due to $O_2$, $H_2O$, $CO_2$ and $CH_4$ and also by the individual gas within the selected wavelength bands of interest. A GENSPECT synthetic line by line radiative transfer model is applied to perform radiative transfer simulations to calculate the radiative flux by varying surface albedo, mixing ratios of the selected greenhouse gases, surface temperature, solar sun and zenith angles with different latitude and longitude of the instrument. Finally, the $SWupRF_{syn}$ estimated from GENSPECT was compared with $SWupRF_{obs}$ from Argus 1000 over a period of four years (2009 and 2013) covering all seasons. We calculate and compare both the synthetic and real measured observed data set. The synthetic model gives $SWupRF_{syn}$ within the range of [0.3950 to 1.650] W/m$^2$ and the selected Argus observed model gives SWupRF$_{obs}$ within the range of [0.10 to 3.15] W/m$^2$. The simulated range of synthetic model also represents 1 to 32 K rise of temperature within the different concentration of water vapor and other gases. The authors determined that the satellite observed data of week 75 pass 43 & week 08 pass 61 showing minimum 0.850 [0.10 to 1.60] W/m$^2$ and maximum 1.65 [0.12 to 3.15] W/m$^2$ respectively.

## 1. Introduction

Knowledge and monitoring of the earth radiation budget is essential for improving our understanding of the Earth climate and potential climate change [1]. The surface radiation flux is a major component of the energy exchange between the atmosphere and the land or ocean surface, and hence affects temperature fields, atmosphere and oceanic circulations [2]. The net surface radiation controls the energy and water exchange between the biosphere and the atmosphere, and has major influence on Earth weather and climate [3]. Earth climate is determined by the flows of energy into and out of the planet and to from earth surface [3, 4]. Changes to the surface energy balance also ultimately control how this hydraulic cycle corresponds to the small energy imbalances that force climate change [4]. Short wave radiation is the energy source of the earth. It plays a key role in the fields of hydrology, meteorology, agriculture, and even climate change [5]. Net surface radiation is the driving force for the surface energy balance and the transportation and exchange of all matters at the interface between the surface and the atmosphere, the net surface shortwave (SW) radiation significantly affects the climate forming and the change of climate [6]. The ability to better monitor each of the shortwave and long wave radiative components at the surface is essential to better understand existing feedbacks between the surface energy and hydrological cycles, and to better assess future effects of climate change. The Earth's surface net shortwave (SW) radiation, *i.e.*, the difference between the incoming and outgoing SW radiation, represents the amount of solar radiation absorbed by the surface and can be derived from satellite observations. [7]. Clouds are the main factor in modulating the Earth energy budget and the Climate [8]. The need of the accurate characterization of SW radiation levels at surface is also very important criteria [8, 9, 10, 11]. In general the traditional ground based radiation measurement can provide an accurate radiation slots, but spatial coverage over the globe seriously limits this application over specially large areas. In comparison with satellite based measurements have a unique advantage to calculate radiative flux over the different areas of globe in terms of special coverage. In early 1970's, NASA has recognized the importance of improving our understanding of the earth radiation budget (ERB), the first Earth Radiation Budget Experiment (ERBE) instrument onboard the Earth Radiation Budget Satellite (ERBS) in 1984 [12,13] and following the clouds and the Earth Radiation Energy System (CERES) [14].

Bulk of the methods providing global coverage of the surface radiation are based on satellite remote sensing. Prior to the advent of the Earth Observation from Space (EOS) era in late 1990s, retrieval schemes employed either the geostationary satellites [15, 16, 17, 18, 19, 20, 21] or the National Oceanic and Atmospheric Administration (NOAA)—operated suite of polar orbiters [22, 23]. Methods used ranged from statistical/empirical schemes [15] to physically-based [16, 17, 18, 19, 20].

Now a days, more and more studies over satellite data are aided to calculate surface shortwave radiative flux [24, 25, 26]. However, all of the above mentioned calculations are good enough for finding the shortwave radiative flux over a large or full wavelength bands. Argus 1000-a micro spectrometer working within NIR region of 1100 to 1700 nm wavelength band along with GENSPECT line by line radiative transfer model [27, 28, 29, 30], shown also a potential ability to accurately calculate the SWupRF within NIR range for the selected greenhouse gases $O_2$, $H_2O$, $CO_2$ and $CH_4$ for the total as well as for the individual wavelength bands with a large spatial area over different locations over the globe. Thus, this study demonstrates that the remote sensing within a small SWupRF wavelength band can be a better foundation for future source for estimating the total energy within the selected NIR wavelength band as well as good source for the efficient detection of cloud scene with background atmospheric profile of temperature and gases mixing ratio concentrations.

## 2.    Instrument and model

### 2.1    Argus 1000- a micro spectrometer.

The Argus 1000 micro-spectrometer as shown in Fig.1, developed at York University, Canada in association with Thoth Technology Inc., is a part of the CanX-2 satellite's payload [31] launched in 2008. CanX-2 orbits in a low Earth orbit (LEO), 640 km above the Earth's surface where Argus's field of view (FOV) provides a spatial resolution of 1.5 km as illustrated in Fig. 2. The Argus 1000 micro-spectrometer operates in the near infrared (NIR) region from 900 to 1700 nm with spectral resolution of 6 nm [27]. The Argus instrument provides a means to make measurements of upwelling radiation reflected to space by the Earth and atmosphere [27]. Reflection spectra of sunlight from the Earth's surface contain significant absorption features associated with the molecular absorption of radiation by particular gas species that can be used to infer the composition of the intervening atmosphere [29, 34]. Argus 1000 records the NIR signature of the surface-troposphere amounts of the significant greenhouse gases Oxygen $O_2$, carbon dioxide (CO2) and water vapour (H2O) in order to monitor anthropogenic pollution and to identify significant sources and sinks in the atmosphere [28, 29]. Methane (CH4), Nitrous oxide (N2O), carbon monoxide (CO) and hydrogen fluoride (HF) species also have absorption features in this spectral region of 1100 nm to 1700 nm [34]. The Technical Specifications of Argus 1000 Spectrometer is shown in Table 1.

Table 1: Technical Specifications, Argus 1000 Spectrometer [32]

| Argus 1000 | Specification |
|---|---|
| Type | Grating spectrometer |
| Configuration | Single aperture spectrometer |
| Field of View | 0.15º viewing angle around centered camera bore sight with 15mm fore-optics |
| Mass | >230 g |
| Accommodation | 45 mm x 50 mm x 80 mm |
| Operating Temp. | -20ºC to +40ºC operating temperature |
| Survival Temp. | -25ºC to + 50ºC survival temperature |
| Detector | 256 element InGaAs diode arrays with Peltier cooler (100 active channels) |
| Grating | 300 g/mm |
| Electronics | microprocessor controlled 10-bit ADC with co-adding feature to enhance precision to 13-bit, 3.6-4.2V input rail 250mA-1500mA (375mA typical) |
| Operational Modes | –Continuous cycle, constant integration time with co-adding feature<br>–Adaptive Exposure mode |
| Data Delivery | Fixed length parity striped packets of single or co-added spectra with sequence number, temperature, array temperature and operating parameters |
| Interface | Prime and redundant serial interfaces RS232 protocol |
| Spectral Channels | 100 (typical) |
| Integration Time | 500 µs to 4.096 sec |
| Handling | Shipped by courier in ruggedized carrying case |

The instrument was designed to take nadir observations of reflected sunlight from Earth's surface and atmosphere. The nadir viewing geometry of Argus is of particular utility as this observation mode provides the highest spatial resolution on the bright land surfaces and returns more useable soundings in regions that are partially cloudy or have significant surface topography [33].

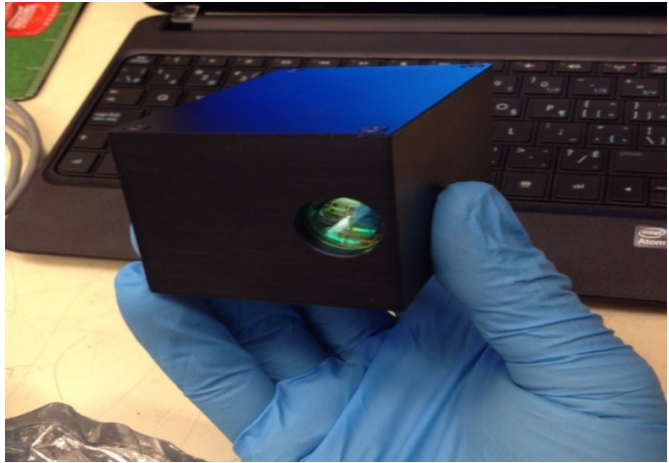

**Figure 1**. Argus 1000 spectrometer at the Space Engineering Laboratory, York University.

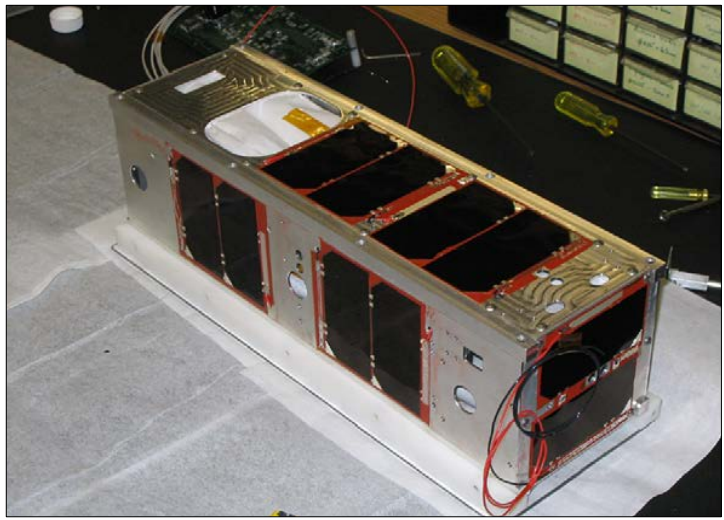

**Figure 2.** Partly integrated CanX-2 spacecraft [Courtesy: UTIAS].

Argus team at the Space Engineering Laboratory at York University prepares the observation tables for the desired targets around the globe using the Systems Tool Kits (STK) software. The Argus-1000 target list contains 35 sites around the Earth. In the last seven years that Argus has been in operation, we have made over 300 observations over a series of land and ocean targets [28], few examples of the observed dataset are as shown in Table 2.

Table 2. Typical Argus week per pass parameters for SWupRF_obs

| Week No._Pass No. | Date | Observation Numbers (OBS) | Packet Length | Observations number with satellite Sun angle, Nadir angle, Lat. & Long. | Location |
|---|---|---|---|---|---|
| Week08_Pass61 | 2009October30 | 32-121 | 276 | OBS32: Sat. nadir angle = 6.2240<br>Sat. sun angle = 35.4821<br>Lat. = 8.6388, Long. = 61.2715<br>OBS121: Sat. nadir angle = 4.9744<br>Sat. sun angle = 33.0354<br>Lat. = 3.0414, Long. = 60.2615 | Arabian Sea & Seychelles |
| Week09_Pass36 | 2009November04 | 22-45 | 69 | OBS22: Sat. nadir angle = 5.0968<br>Sat. sun angle = 35.5000<br>Lat. = 6.8350, Long. = -35.4025<br>OBS45: Sat. nadir angle = 5.4026<br>Sat. sun angle = 33.9916<br>Lat. = 3.7837, Long. = -36.0342 | North Atlantic Ocean |
| Week14_Pass52 | 2010March04 | 22-125 | 125 | OBS22: Sat. nadir angle = 24.7152<br>Sat. sun angle = 58.1136<br>Lat. = 47.3740, Long. = -77.7286<br>OBS125: Sat nadir angle = 21.8777<br>Sat. sun angle = 54.5434<br>Lat. = 42.2893, Long. = -79.9453 | Toronto/ Kitcisakik (Canada) |
| Week14_Pass54 | 2010March04 | 14-123 | 123 | OBS14: Sat. nadir angle = 4.1527<br>Sat. sun angle = 62.7967<br>Lat. = 53.1442, Long. = -124.6127<br>OBS123: Sat nadir angle = 17.2101<br>Sat. sun angle = 57.9201<br>Lat. = 47.0893, Long. = -126.6271 | Vancouver (Canada) |
| Week17_Pass42 | 2010April28 | 14-144 | 144 | OBS14: Sat. nadir angle = 18.2065<br>Sat. sun angle = 51.0290<br>Lat. = 64.8690, Long. = 177.4778<br>OBS144: Sat nadir angle = 1.2381<br>Sat. sun angle = 43.9990<br>Lat. = 55.5832, Long. = 172.3671 | Magadan (Russia) |
| Week30_Pass46 | 2010December 16 | 35-460 | 460 | OBS35: Sat. nadir angle = 2.6513<br>Sat. sun angle = 32.3060<br>Lat. = -26.7341, Long. = 43.2740<br>OBS460: Sat nadir angle = 2.2886<br>Sat. sun angle = 45.0506<br>Lat. = -55.4500, Long. = 33.7905 | Indian Ocean |
| Week41_Pass27 | 2011September08 | 14-198 | 198 | OBS14: Sat. nadir angle = 4.5387<br>Sat. sun angle = 56.8632<br>Lat. = -37.5321, Long. = -72.0149<br>OBS198: Sat nadir angle = 3.0990<br>Sat. sun angle = 67.1288<br>Lat. = -50.2686, Long. = -76.2329 | Patagonia (South America) |
| Week75_Pass43 | 2013August14 | 19-65 | 65 | OBS19: Sat. nadir angle = 1.8888<br>Sat. sun angle = 38.6809<br>Lat. = 31.7593, Long. = 148.2136<br>OBS65: Sat nadir angle = 1.6877<br>Sat. sun angle = 38.1652<br>Lat. = 19.9247, Long. = 145.5481 | North Pacific Ocean |

## 2.2 GENSPECT – a line by line radiative transfer model

The GENSPECT is a line-by-line radiative transfer algorithm for absorption, emission, and transmission for a wide range of atmospheric gases [29]. Given information including gas types and amounts, pressure, path length, temperature, and frequency range for an atmosphere, the GENSPECT model computes the spectral characteristics of the gas. GENSPECT employs a new computation algorithm that maintains a specified accuracy for the calculation as a whole by pre-computing where a line function may be interpolated without a reduction in accuracy. The approach employs a binary division of the spectral range, and calculations are performed on a cascaded series of wavelength grids, each with approximately twice the spectral resolution of the previous one. The GENSPECT error tolerances are 0.01%, 0.1%, and 1%, which may be selected according to the application [29]. GENSPECT is used in this study to simulate the real atmosphere by dividing it into plane-parallel layers to handle the vertical stratification of the atmosphere.

## 3. Data sets with model simulations

For modeling the SWupRF over different locations, the different weeks per passes of Argus data set is applied as shown in Table 2. In this study, the GNESPECT model along with Argus 1000 is applied by selecting the sun elevation angle, satellite nadir angle, variable path length, atmospheric water vapor, variable albedo, and with or without cloud structure over land or sea. Table 3 have shown the few input parameters used for calculating SWupRF. Fig. 3 & 4 shows the GENSPECT-Synthetic spectrum with different H2O concentrations and albedo, & satellite sun and nadir angle of the selected Argus week per pass per observation numbers. Fig 4 also shows the dominant increase of radiance shift by changing water vapor concentration, surface albedo and altitudes from surface to reflecting medium. Both the spectrum shown a good absorption features of $O_2$ (at 1260 nm), $H_2O$ (at 1200 nm &1400 nm), $CO_2$ (at 1570 nm & 1610 nm) & CH$_4$ (at 1670 nm). The selected wavelength and other parameters for GNESPECT synthetic model is shown in Table 4.

Table 3. Input parameters for SWupRF$_{syn}$ model.

| Types of parameter | Significance values and ranges |
|---|---|
| Mixing Ratios of gases | refmod 95_ O$_2$.mxr, refmod 95_ CO$_2$.mxr, refmod 95_ CH$_4$.mxr, refmod 95_ H$_2$O.mxr (1976 U.S. Standard Atmospheric Model) |
| Gases in % | O$_2$ (100) , CO$_2$ (100), CH$_4$ (100), H$_2$O (0 to 35) |
| Height from surface to top of clouds | 2km to 50 km |
| Surface Type | Lambertian |
| Surface Temperature | 288$^0$K to 300$^0$K |
| Satellite sun angle* | Argus geo location (Obs no.) |
| Satellite nadir angle* | Argus geo location (Obs no.) |
| Reflectivity | 0.3 (over generic vegetation and bare soil)<br>0.1 to 0.9  (over snow, clouds, and ice) |
| Scattering Type | Rayleigh |

*Different for each Argus week/pass/observation number applied for calculating SWupRF

Table 4: Wavelength and smoothing bands used for GENSPECT synthetic model

| Name of Slices | Wavelength Bands Range (nm) | Smoothening Bands | Slope Threshold* |
|---|---|---|---|
| O$_2$  absorption band | [1150 - 1330] | O$_2$  synthetic,<br><br>O$_2$  observed [50,50] | 0.5e-6 |
| H$_2$O absorption band | [1340 - 1440] | H$_2$O synthetic, H$_2$O observed [50,50] | 0 |
| CO$_2$ absorption band | [1580 - 1620] | CO$_2$ synthetic, CO$_2$ observed [0,0] | 0 |
| CH$_4$ absorption band (max 1675!!) | [1625 - 1673] | CH$_4$  synthetic, CH$_4$ observed [0,0] | 0 |

*Plateau detection threshold for the O$_2$ band (its variable height and % position requires us to detect the sub-band where it is most "flat").

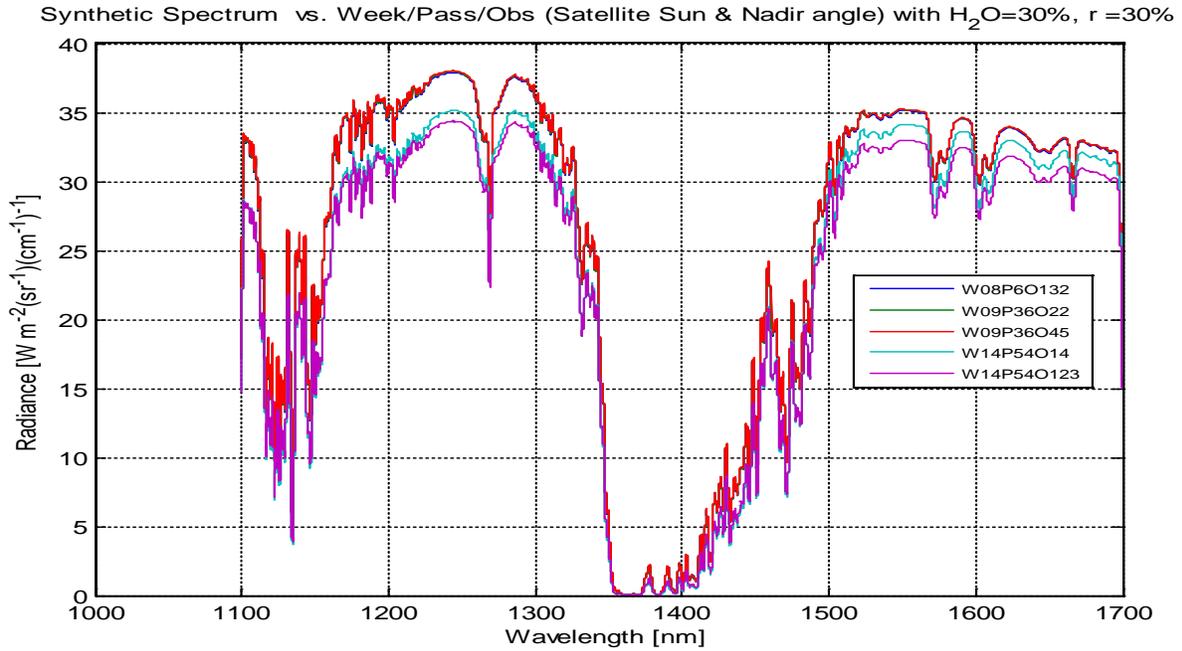

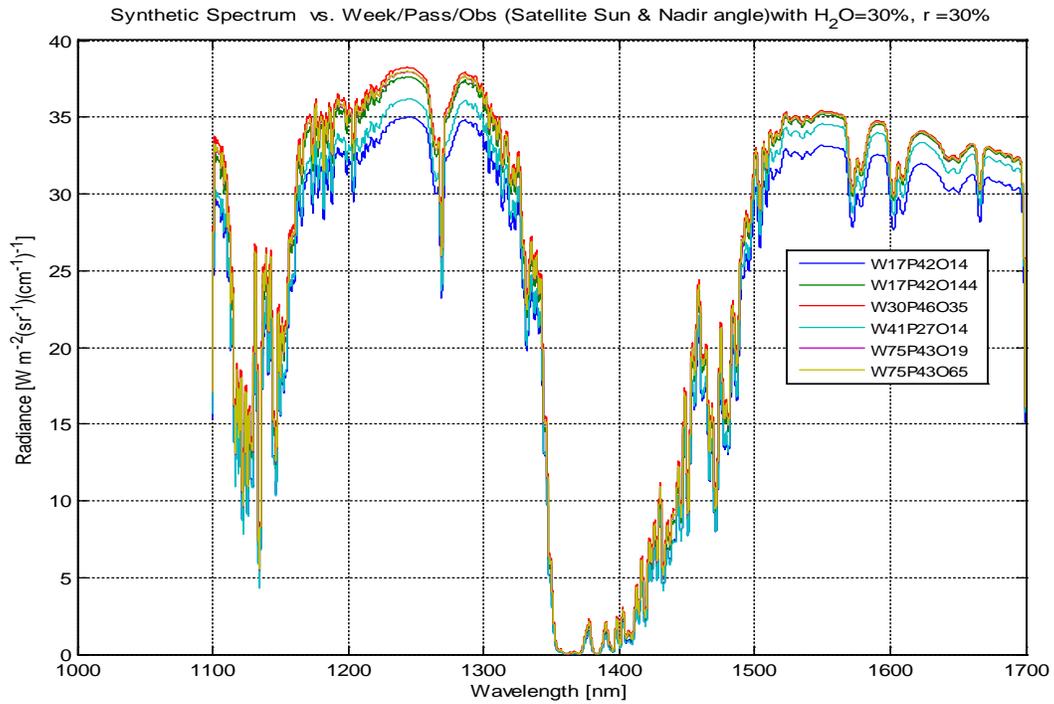

Fig. 3 (a) & (b) : GENSPECT-Synthetic spectrum with H2O = 30%, albedo = 0.3, by varying satellite sun and nadir angle for selected week per pass per observation number.

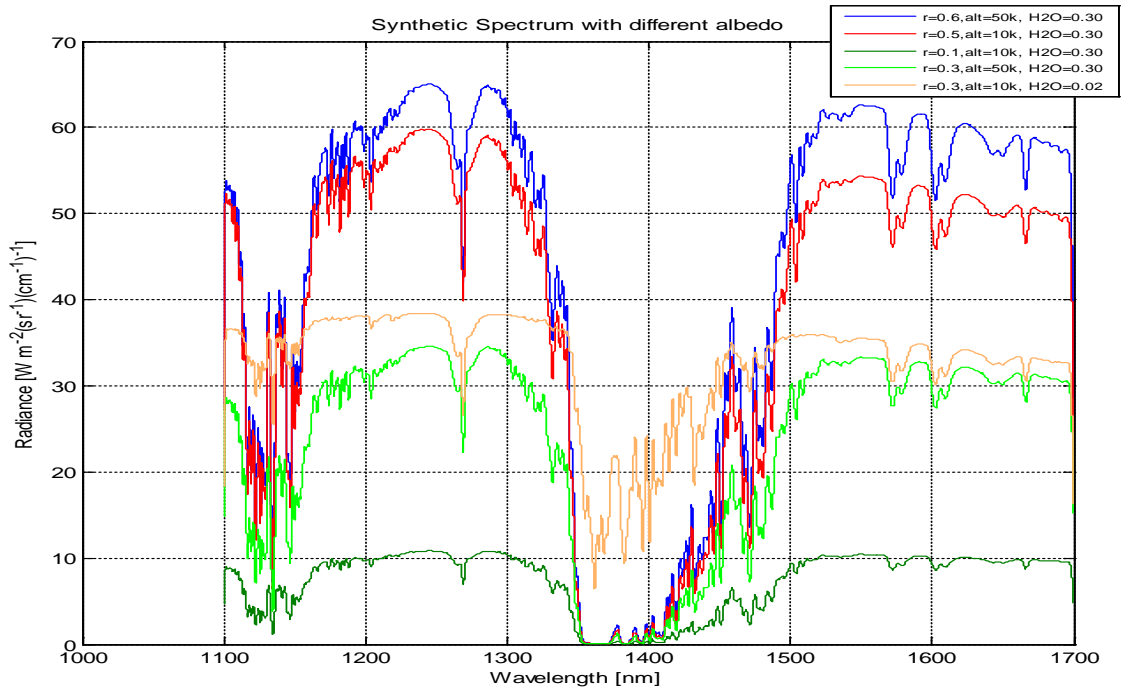

Fig. 4: GENSPECT-Synthetic spectrum with different albedo, altitudes and concentration of $H_2O$.

The Argus-1000 target list contains 35 sites around the Earth. In the last seven years that Argus has been in operation, we have made over 300 observations over a series of land and ocean targets [28]. Moreover, Argus can also detect a cloud scene within the selected range of week per passes of Argus flight [28].

In order to validate the Argus space data results, a radiative transfer simulations were again carried out by GENSPECT synthetic model included the actual solar sun and zenith angle of the space instrument as shown in table 1. The Argus spectrum profile of different selected weeks per passes with observation numbers as shown in Fig. 5 and Fig. 6. Most of the spectrums in Fig. 5 & Fig. 6 have shown a good absorption features of $O_2$, $H_2O$, $CO_2$ & $CH_4$ within the specified wavelength bands, which is used for the calculation of SWupRF for both synthetic and real observations, different scenrios were simulated within the instrument range.

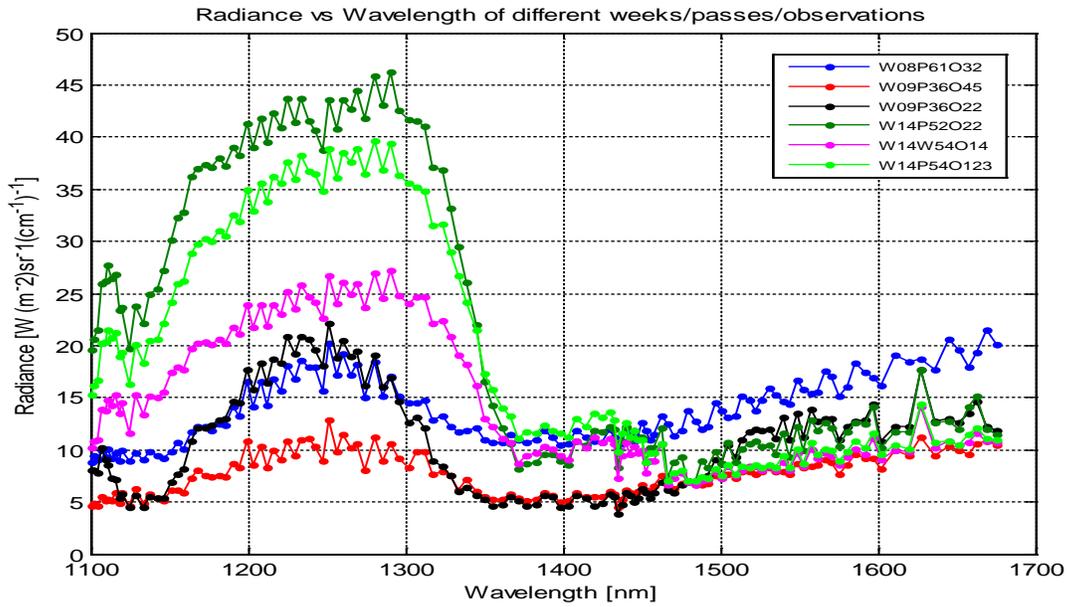

Fig. 5: Argus spectra- radiance vs wavelength of weeks per passes with selected observation numbers (week08/09/14, pass61/36/52/54, obs32/45/22/14/123).

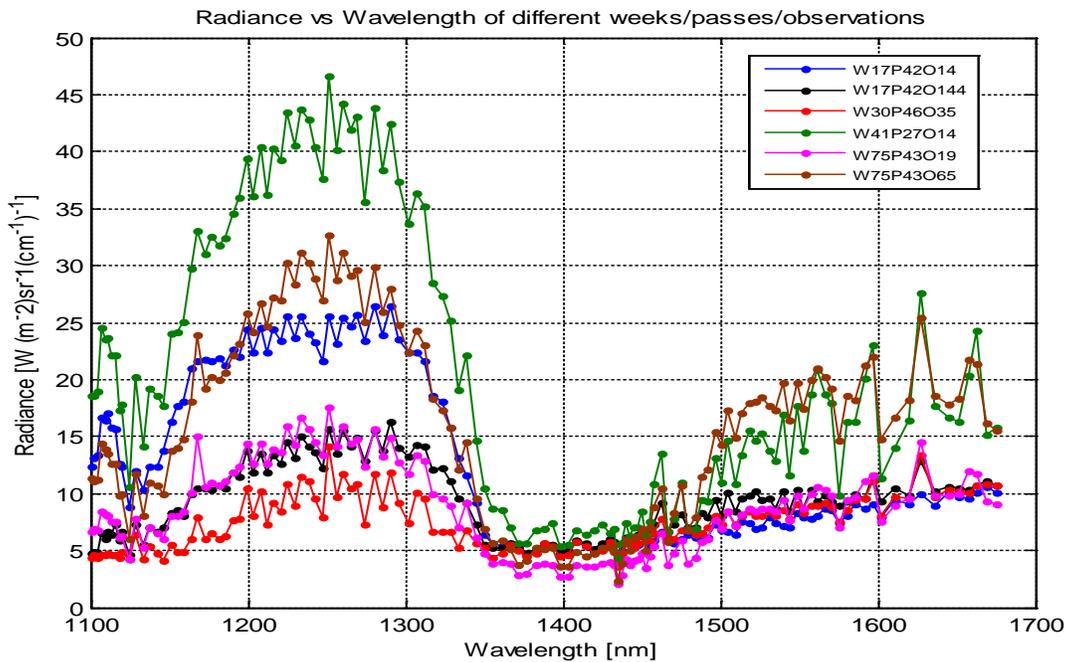

Fig. 6: Argus spectra- wavelength vs radiance of different weeks per passes with selected observation numbers week17/30/41/75, pass42/46/27/43, obs14/144/35/19/65).

.

## 4.    Methodology

Any retrieval method is likely to be loaded with difficulty as no method can completely capture the complex behavior of the atmosphere. Even if the observational problems can be overcome the potential utility of such data may not be clear [35]. We used the method of optimal retrievals by means of a forward model to output simulated radiance observations. The spectral distribution of solar radiation, or extraterrestrial solar spectrum, incident on the top of the Earth's atmosphere is shown in Fig. 7. The interaction of this radiation with the surface and atmosphere is the source of the reflected radiation measured by Argus. Absorption and scattering must be accounted for along the incoming and outgoing atmospheric paths that depend on the sun-Earth-satellite geometry includes solar zenith angle, satellite viewing angle and relative azimuth with respect to the point of observation.

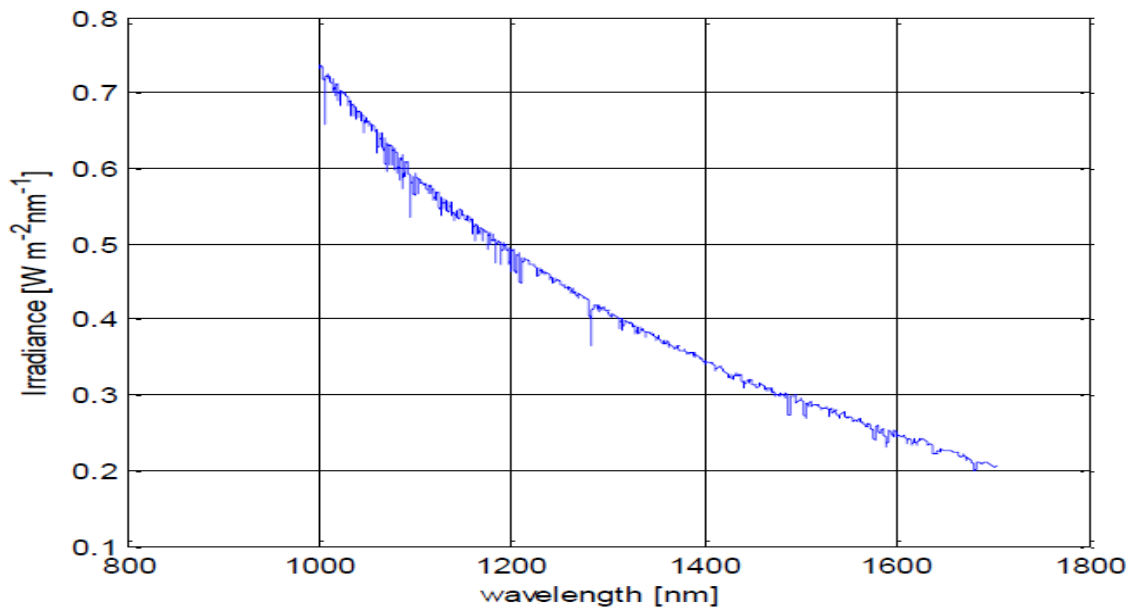

Fig 7. Solar Spectrum for Argus spectral window Reproduced from the 2000 ASTM Standard
        Extraterrestrial Spectrum Reference E-490-00. [34]

The SWupRF model have been developed by the retrievals spectra of Argus 1000 alongwith GENSPECT line by line radiative transfer model to find the total intensity of each Argus spectra by integrated absorption technique using eq. no. (1) to (4) which integrates over the full spectral range.

$$(SWupRF)syn, (SWupRF)obs = \int_{\lambda 1}^{\lambda max} S\,(\lambda)d\lambda \tag{1}$$

where $(SWupRF)syn$ = Short Wave upwelling Radiative Flux of GENSPECT synthetic spectra

$(SWupRF)obs$ = Short Wave upwelling Radiative Flux of Argus observed spectra

$S\,(\lambda)$ = Spectral radiance [(Wm$^{-2}$ sr$^{-1}$ (1/cm)$^{-1}$]

$\lambda 1\ to\ \lambda max$ = wavelength band

The equation (1) can also be written as,

$$(SWupRF) \cong \sum_{i=1}^{N} S\,(\lambda i)d\lambda \tag{2}$$

To integrate over a solid angle the following equations is applied:

$$((SWupRF)_\Omega = (SWupRF) \times 2\pi \int_0^{\frac{\pi}{2}} sin\theta cos\theta d\theta = \pi(SWupRF) \tag{3}$$

The total spectral SWupRF is calculated by the following relationship:

$$\mathbf{(SWupRF)_{total}} = \mathbf{(SWupRF)_\Omega} \times r_v \tag{4}$$

where $r_v = resolution\ wavenumber$

By applying the above relationship the total SWupRF$_{syn}$ for a selected greenhouse gases was calculated with each wavelength band by simulating the upwelling surface flux within the NIR range of instrument and recalculating the SWupRF with the range of different percentages of $H_2O$ and $CO_2$. The mixing ratios concentration of $O_2$ and $CH_4$ were held constant (US1976 atmospheric model) throughout the atmospheric profile from 2 to 50 km. The temperature profile were varied from $288^\circ$ K to $300^\circ$ K. The SWupRF$_{obs}$ were calculated within the wavelength range of weeks per passes of Argus 1000 by choosing solar sun and zenith angle of the instrument observation as shown in table 1.

## 5.    Result and discussions:

### 5.1    SWupRF$_{syn}$ by GENSPECT radiative transfer model:

Fig. 8 (a), (b), (c) & (d) illustrate the spectral response of GENSPECT synthetic output with the different concentration of water vapor ranges from 2% to 45%. All the generated spectra clearly indicate the strong features of absorption of $O_2$, $H_2O$, $CO_2$ and $CH_4$. This can easily be exemplified from the spectral trend within the water vapor band that it may become spectrally saturated while reaching around 35% of mixing ratio of $H_2O$ concentration.

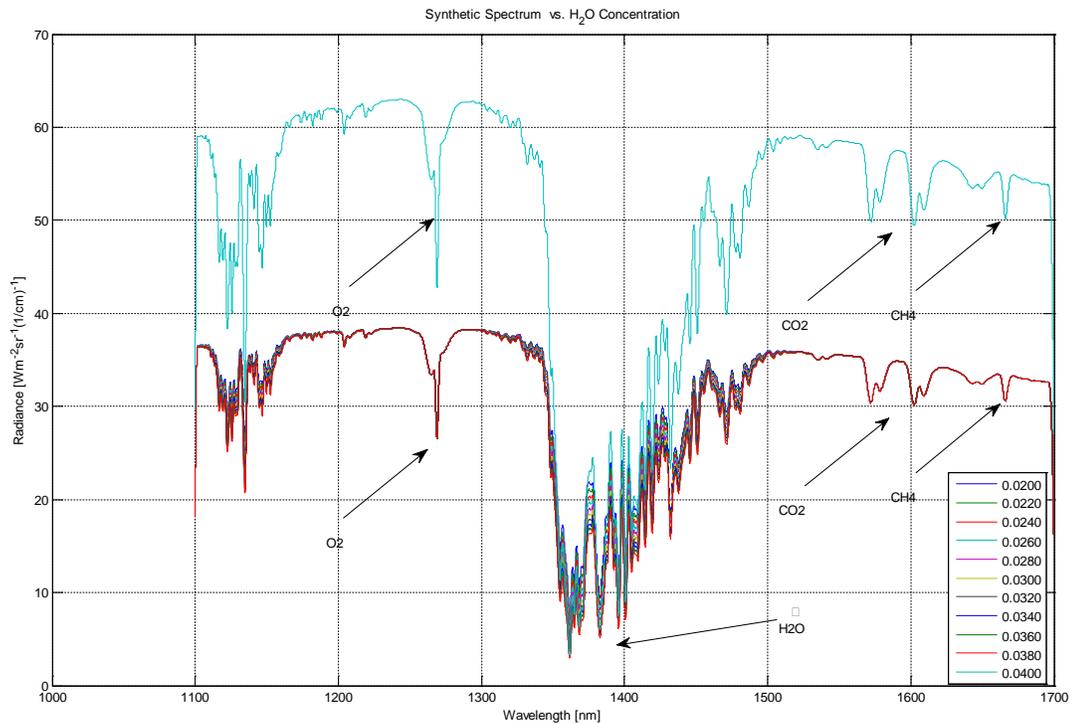

Fig.8 (a) Synthetic spectra of $H_2O$ from 2% to 4% of original atmospheric model

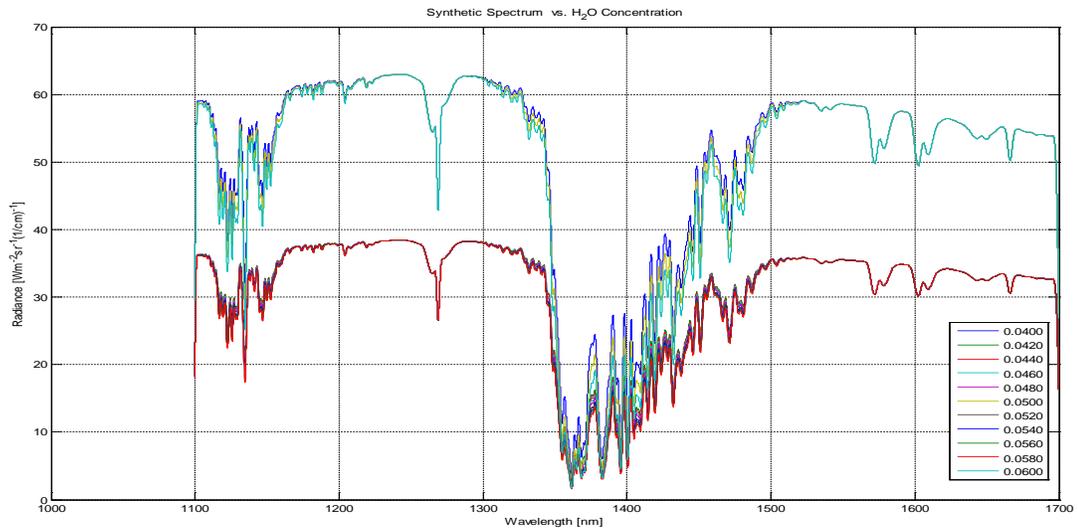

Fig.8 (b) Synthetic spectra of $H_2O$ from 4% to 6% of original atmospheric model

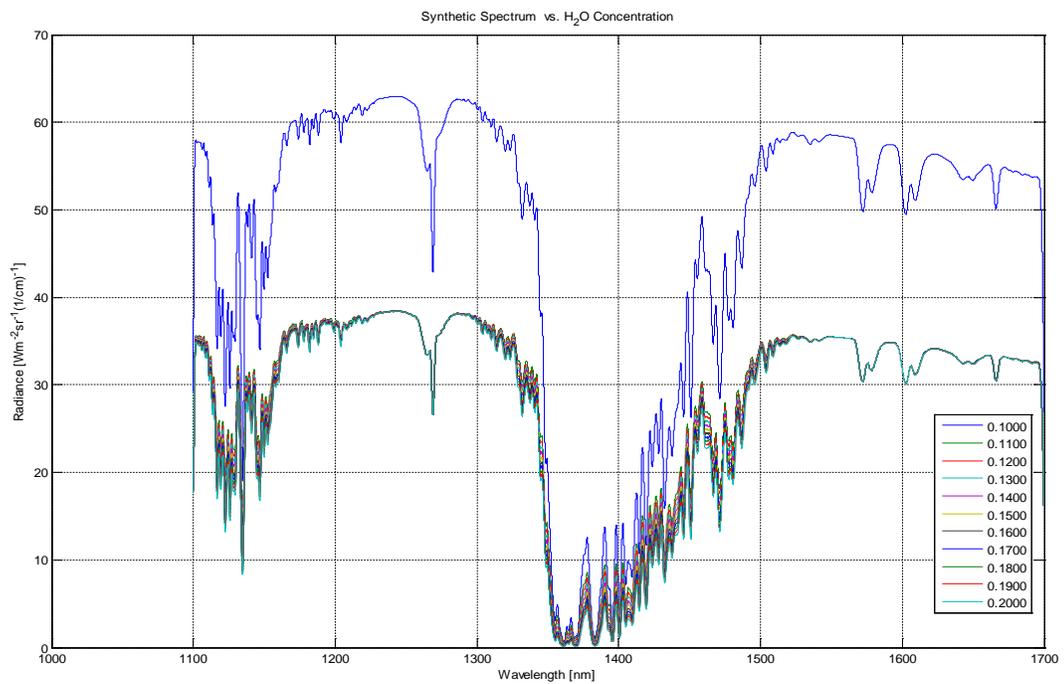

Fig.8 (c) Synthetic spectra of $H_2O$ from 10% to 20% of original atmospheric model

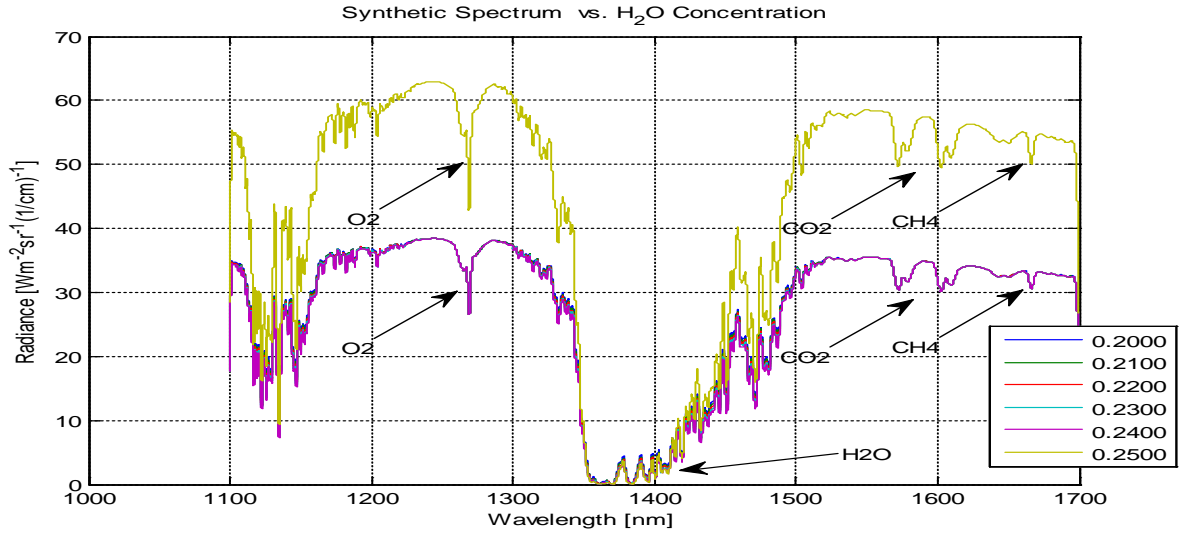

Fig.8 (d) Synthetic spectra of $H_2O$ from 20% to 25% of original atmospheric model

Different synthetic spectra of $CO_2$ have also been generated by fixing water vapor concentration (same as in US atmospheric model and also at 35% of $H_2O$ while saturated) and changing $CO_2$ concentration from 0% to 200% as shown in Fig.9.

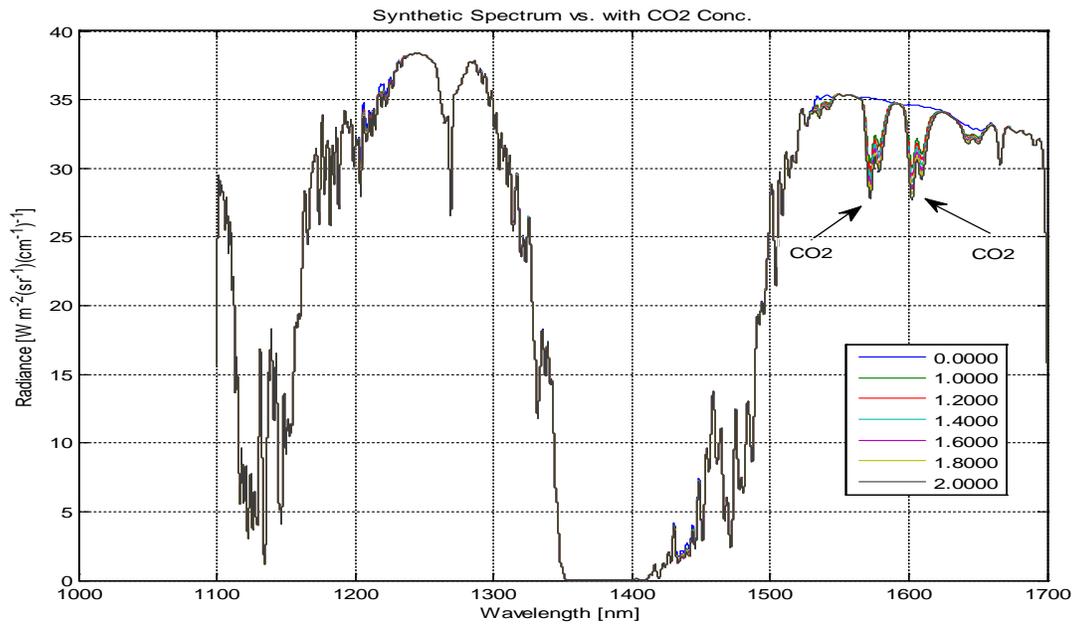

Fig.9 Synthetic spectra of $CO_2$ from 0% to 100% and $2CO_2$ of original atmospheric model

All the simulated spectra have been used to calculate the $(SWupRF)_{syn}$ by applying the Integrate synthetic spectra model illustrated in Fig 10 & Fig 11, clearly demonstrate the total short upwelling wave radiative flux within the 1100 nm to 1700 nm of wavelength bands due to the overall effect of $O_2$, $H_2O$, $CO_2$ and $CH_4$ while integrating over total wavelength bands. The trends clearly indicate that the radiative flux intensity in decreasing while increasing the concentration of water vapor within the selected wavelength bands of interest. This shows that the short wave upwelling radiative flux is very sensitive to $H_2O$ when the water vapor concentration is high.

Fig 10 (a) & (b) exemplify the $(SWupRF)_{syn}$ is 0.4725 [0.3950 to 0.5500] W/m$^2$ & 1.340 [1.030 to 1.650] W/m$^2$ with albedo =0.3 & 0.9 respectively, at different surface temperatures, fixed concentration of $O_2$, $CO_2$ & $CH_4$ by varying water vapor.

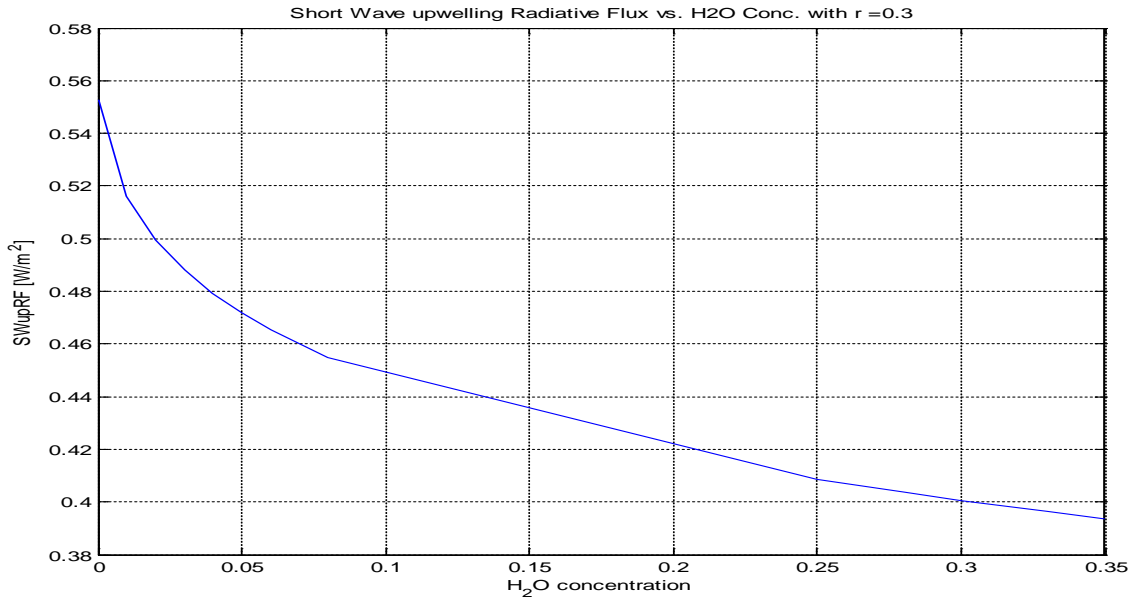

Fig.10 (a) $(SWupRF)_{syn}$ from 0% to 35% of $H_2O$ concentration with albedo = 0.3

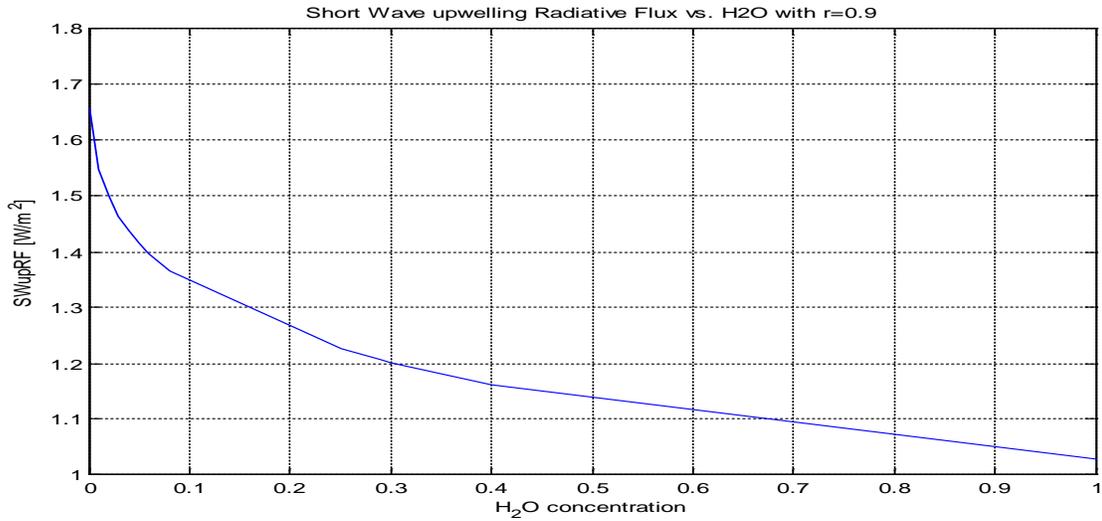

Fig.10 (b) $(SWupRF)_{syn}$ from 0% to 35% of $H_2O$ concentration with albedo = 0.9

The different simulated synthetic spectra of $CO_2$ band from 1565 nm to 1620 nm have given substantial absorption of $CO_2$ in comparison with other spectra from 1% to 200% variation of $CO_2$. The estimate of the $(SWupRF)_{syn}$ with different surface temperature, fixed concentration of $O_2$, $H_2O$ & $CH_4$ with varying carbon dioxide is 0.3430 [0.3407 to 0.3454] W/m².

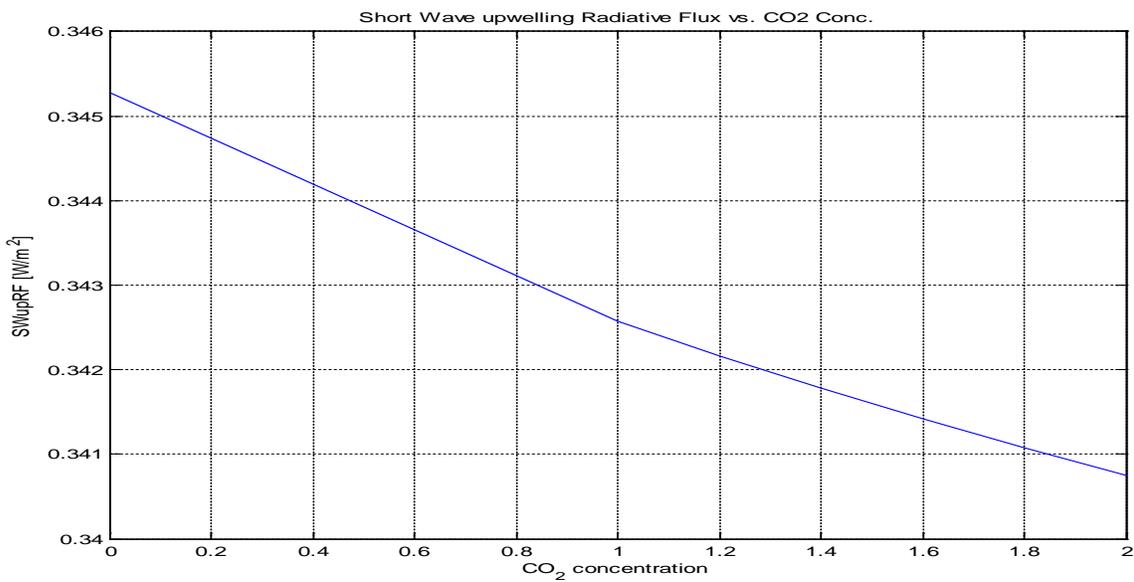

Fig.11 $(SWupRF)_{syn}$ at 0% to 200% of $CO_2$ concentration

## 5.2    SWupRF$_{obs}$ by Argus 1000 space data:

Fig. 12 to Fig 15 illustrated the spectral response of Argus week per pass per observation for the full range of satellite raw data of Argus flight. Similarly in contrast with GENSPECT model, the strong absorption for O$_2$ was noticed at about 1260 nm. We also found the absorption signatures of CO$_2$ at around 1570 nm and 1610 nm and as well as CH$_4$ absorption features at 1670 nm. Fig 13 shown the strong absorption features of H$_2$O within 1390 nm to 1430 nm wavelength band, which tends to become saturate in this spectral region.

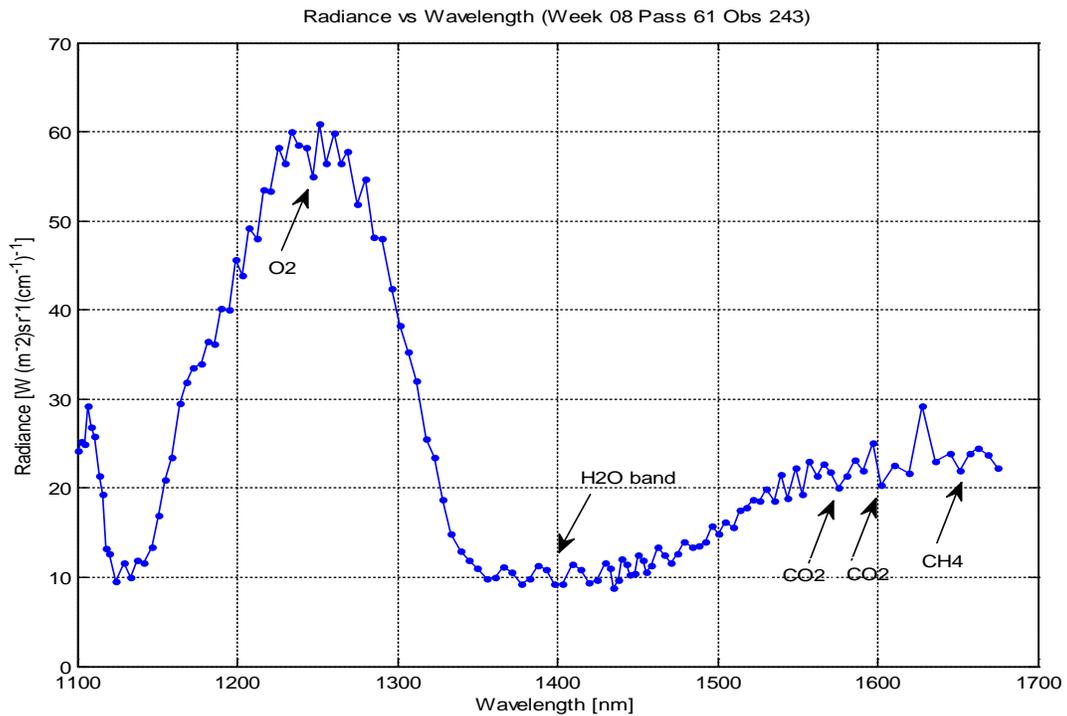

Fig.12 Argus observed spectra of week 08 pass 61 with observation 243

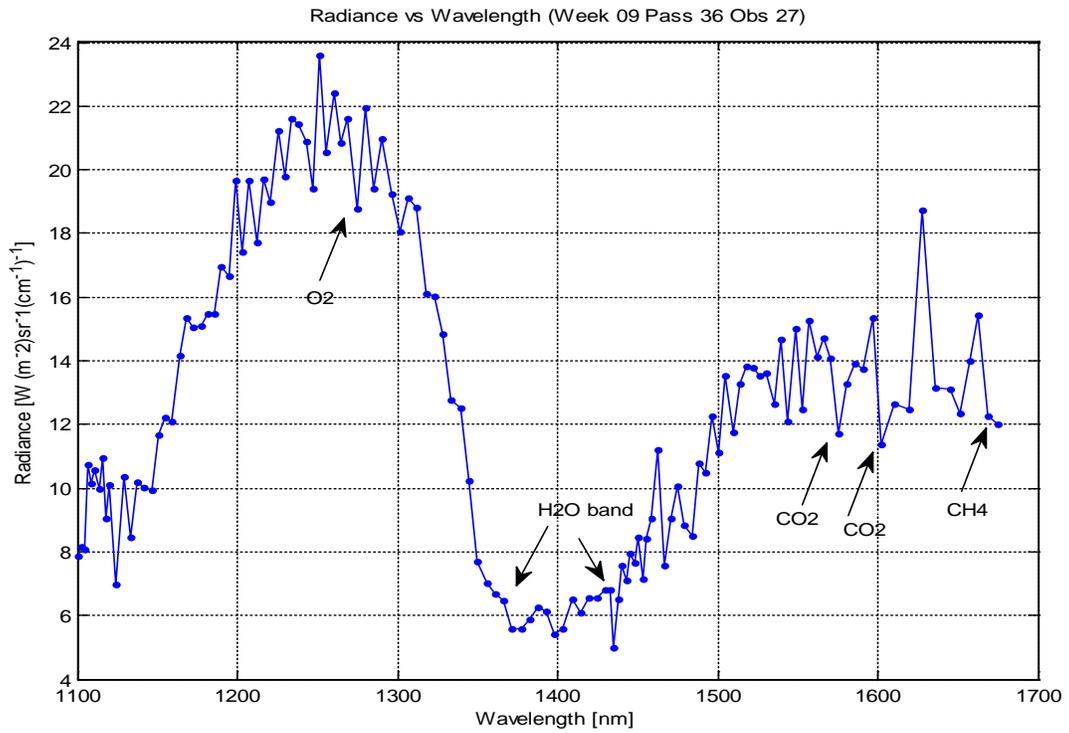

Fig.13 Argus Observed spectra of week 09 pass 36 with observation 27

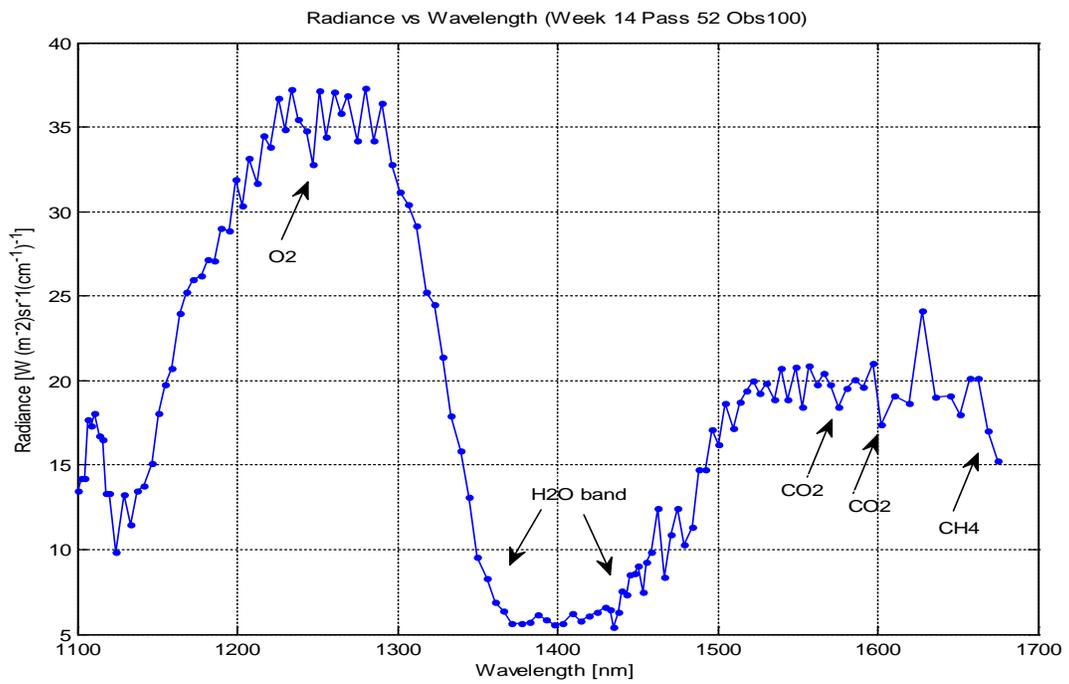

Fig.14 Argus Observed spectra of week 14 pass 52 with observation 100

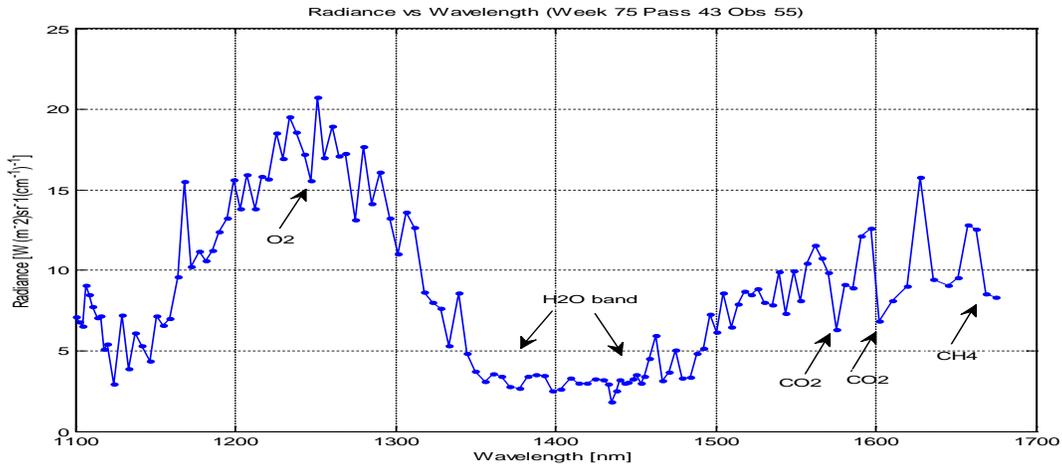

Fig.15 Argus Observed spectra of week 75 pass 43 with observation 56

All the observed spectra with different week per pass number with a set of observations as shown in Table 2 have been used to calculate the $(SWupRF)_{obs}$ by applying the 'Integrate observed spectra model'. Fig. 16 shown a comparison of few with GENSPECT simulated radiance with albedo 0.3 and 0.5, altitude variations from 10 to 50 km and fixed 30 % $H_2O$ concentration with Argus observations profile numbers 20,100,120 and 124 of week 14 pass 52 recorded by on March 4th , 2010 over Vancouver, Canada (see Table 2). This spectrum for Argus 1000 spectrometer shows the absolute radiance value around 6 to 60 $Wm^{-2}sr^{-1}(cm^{-1})^{-1}$.

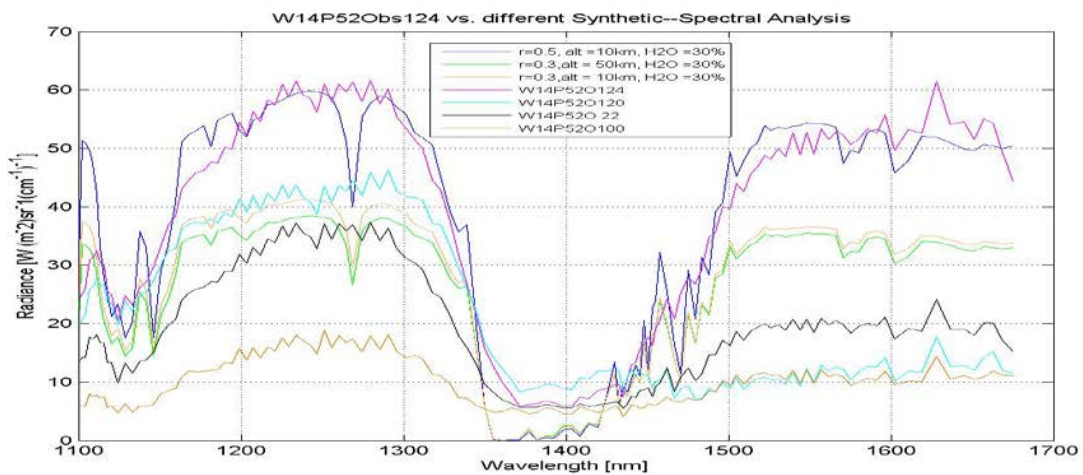

Fig.16 Different Synthetic vs. Argus observed spectra of week 14 pass 52

Model parameters are optimized to achieve best-fit retrieval results as shown in Fig 17. The gases used in model are $O_2$, $H_2O$, $CO_2$ and $CH_4$, Rayleigh single scattering is included in the model. The surface reflectivity is determined iteratively to be 0.3 (assuming a Lambertian surface). The atmospheric composition model used is RefMod 2000 and atmospheric density model is from the US standard Atmosphere. The spacecraft nadir view angle is $4°$ to $17°$ and Sun angle is $57°$ to $62°$ for this selected data set of Argus flight. Table 5 illustrates the integrated output of few selected data set of Argus observed spectra of different weeks per passes per observations.

Table 5: Argus - SWupRF$_{obs}$ (W/m$^2$) model output

| Argus Week/Pass | Date | Geo-location | SWupRF$_{obs}$ (W/m$^2$) |
|---|---|---|---|
| 08/61 | October 30th, 2009 | Arabian Sea & Seychelles | 1.65 [0.12 to 3.15] |
| 09/36 | November 4th, 2009 | North Atlantic Ocean | 1.095 [0.011 to 2.180] |
| 14/52 | March 4th, 2010 | Toronto, Canada | 1.215 [0.13 to 2.30] |
| 17/42 | April 28th, 2010 | Magadan, Russia | 0.935 [0.051 to 1.82] |
| 41/27 | September 8th, 2011 | Patagonia (South America) | 0.915[0.10 to 1.73] |
| 75/43 | August 14th, 2013 | Greenland Sea, North of Iceland | 0.850 [0.10 to 1.60] |

The high values of SWupRF are due the probability of TOA layers of thick clouds, aerosols and dense water concentration and the low values are due the clear sky, thin clouds, directly due to earth surface and oceans.

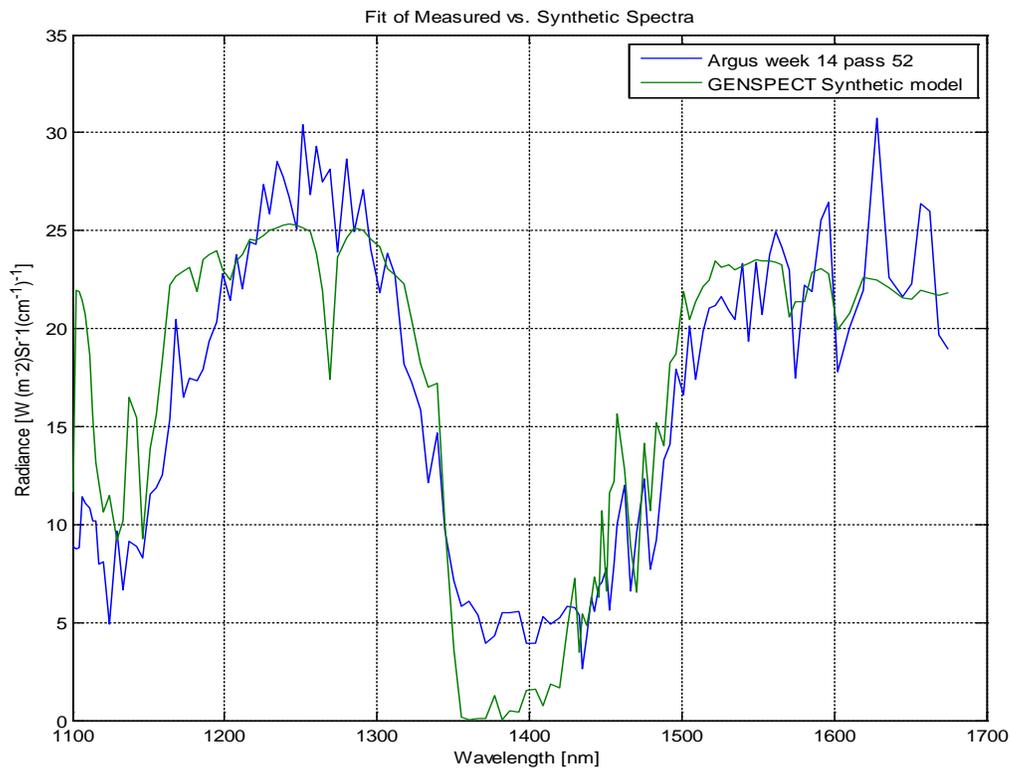

Fig.17  Best fitted synthetic vs. Argus observed spectra of week 14 pass 52.

Fig. 18 illustrates the full set of Argus observed spectra of week 08 pass 61 (October 30th, 2009) over

Arabian Sea & Seychelles. The average range of $(SWupRF - week08pass61)_{obs}$ is 1.65 [0.12 to

3.15] W/m². In this

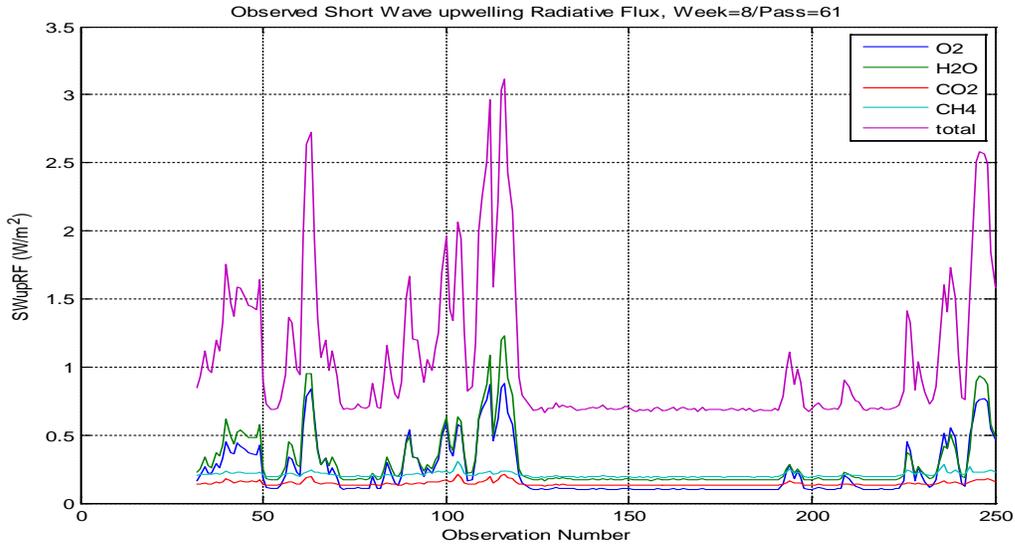

Fig.18. $(SWupRF)_{obs}$ of week 08 pass 61 of Argus observed data

Fig. 19 illustrates the full set of Argus observed spectra of week 09 pass 36 (November 4th, 2009) over North Atlantic Ocean. The average range of $(\boldsymbol{SWupRF-wee09pass36})_{obs}$ is 1.095 [0.011 to 2.180] W/m².

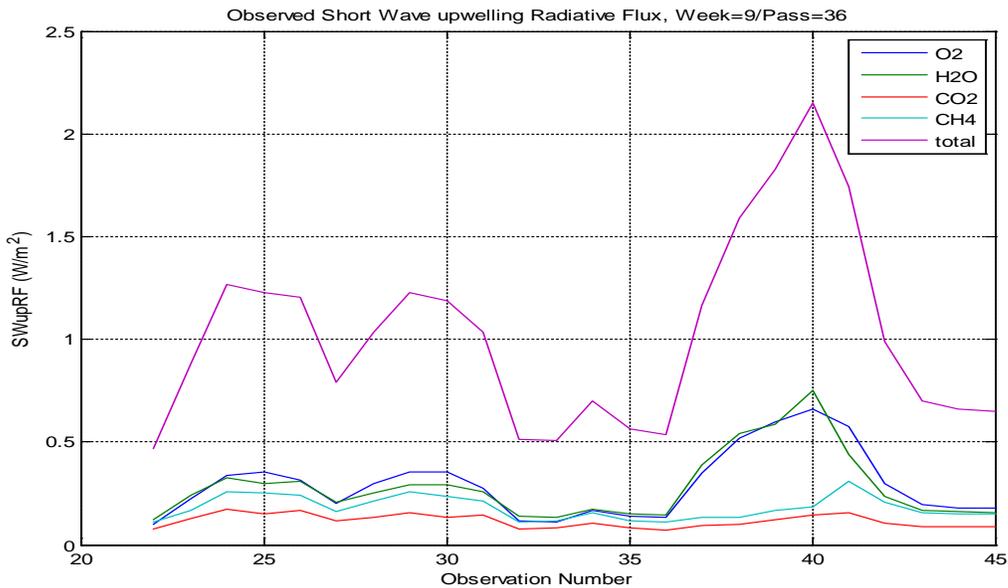

Fig.19. $(SWupRF)_{obs}$ of week 09 pass 36 of Argus observed data

Fig. 20 illustrates the full set of Argus observed spectra of week 14 pass 52 (March 4[th], 2010) over Toronto, Canada. The average range of $(SWupRF - week14pass52)_{obs}$ is 1.215 [0.13 to 2.30] W/m².

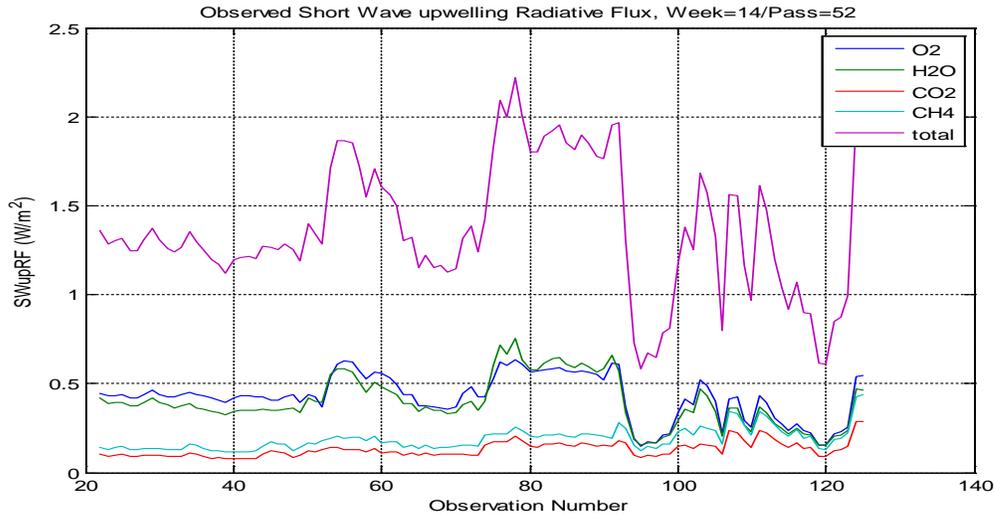

Fig.20. $(SWupRF)_{obs}$ of week 14 pass 52 of Argus observed spectra

Fig. 21 illustrates the full set of Argus observed spectra of week 17 pass 42 (April 28[th], 2010) over Magadan, Russia. The average range of $(SWupRF - week17pass42)_{obs}$ is 0.935 [0.051 to 1.82] W/m².

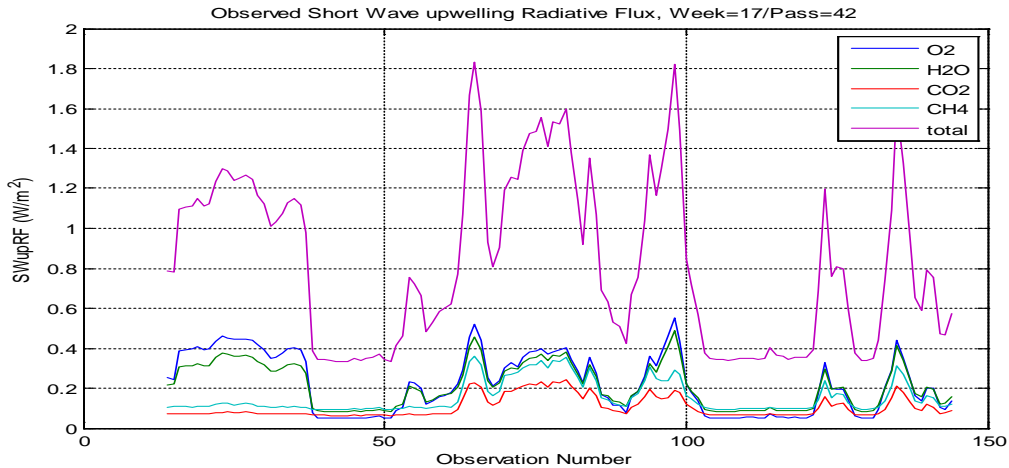

Fig.21. $(SWupRF)_{obs}$ of week 17 pass 42 of Argus observed spectra

Fig. 22 illustrates the full set of Argus observed spectra of week 41 pass 27 (September 8th, 2011) over Patagonia (South America). The average range of $(SWupRF - week41pass27)_{obs}$ is 0.915[0.10 to 1.73] W/m².

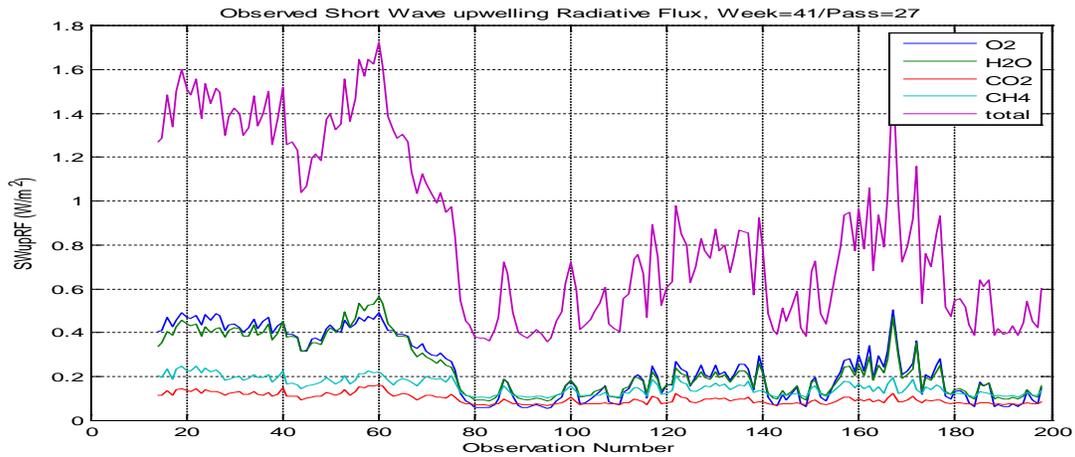

Fig.22. $(SWupRF)_{obs}$ of week 41 pass 27 of Argus observed spectra

Fig. 23 illustrates the full set of Argus observed spectra of week 75 pass 43 (August 14th, 2013) over Greenland Sea, North of Iceland. The average range of **$(SWupRF - week75pass43)_{obs}$** is 0.850 [0.10 to 1.60] W/m².

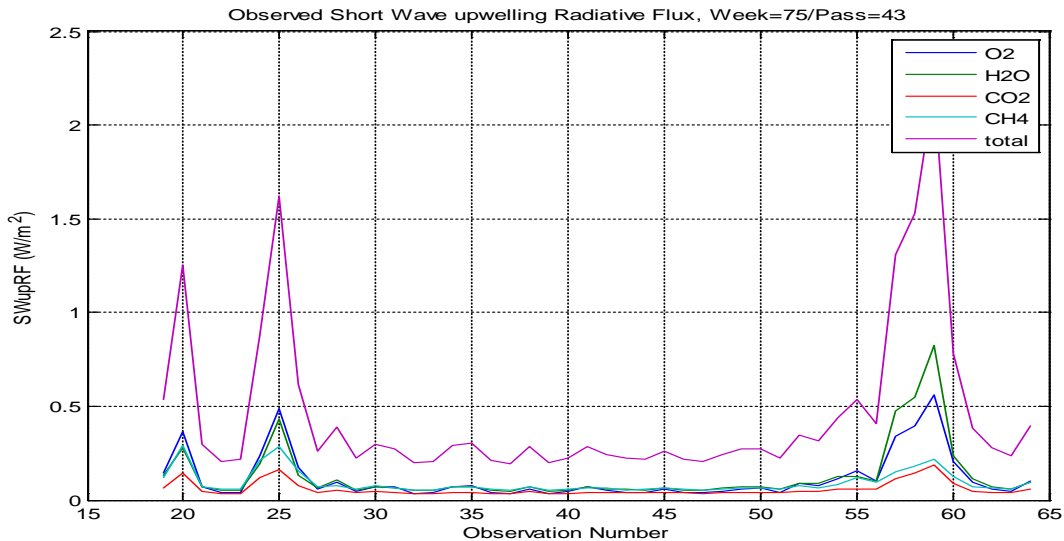

Fig.23. $(SWupRF)_{obs}$ of week 75 pass 43 of Argus observed spectra

From Fig 18 to Fig 23, have given an idea of the total radiative flux for both the synthetic and observed SWupRF within the selected NIR range of Argus 1000 for the selected total and individual bands of $O_2$, $H_2O$, $CO_2$ and $CH_4$ shown different intensity of upwelling radiation. The $O_2$ and $H_2O$ have high radiative flux probably due the high altitude at top of the atmosphere as well as high mixing ratio of water vapor. The weak features of $CO_2$ and $CH_4$ in terms of flux intensity were now incorporated approximately as overlapping absorbers. The high and low radiative fluxes were also due to atmospheric gas concentrations, geolocations of Argus flight (solar sun and zenith angles, latitude and longitudes), different types of earth surfaces, types of clouds and aerosols layers etc. By having an investigation with few outputs of both satellite week per pass per observation of Argus 1000 along with the synthetic-GENSPECT model has been given a clear understanding of the total radiative flux within SWupRF-NIR wavelengths range of terrestrial emission by the selected four greenhouse gases on Earth environments and specially by $H_2O$ and $CO_2$ as they both acts as a most important greenhouse gas and interface significantly with the band of the radiatively active gas in term of Short Wave upwelling Radiative Flux in (W/m$^2$).

## 6.    Conclusions

Surface radiative flux of SW within NIR spectral range from 1100 nm to 1700 nm were determined using the satellite based measurement over different spatial locations during the period of 2009 to 2013 under different atmospheric concentrations. Short Wave upwelling Radiative Flux ($SWupRF$) for $O_2$, $H_2O$, $CO_2$ and $CH_4$ were simulated within the four selected wavelength bands by applying GENSPECT radiative transfer line by line code. The same methodology has also been applied to calculate the ($SWupRF$) of the satellite observed raw data of Argus 1000 spectrometer. The synthetic model gives ($SWupRF$)$_{syn}$ within the range of [0.3950 to 1.650] W/m$^2$ and the selected Argus observed model gives ($SWupRF$)$_{Obs}$ within the range of [0.01 to 3.15] W/m$^2$. The simulated results of ($SWupRF$)$_{syn}$ with the same set of solar sun and zenith angles were compared with the few measured results of $(SWupRF)_{Obs}$ of the Argus satellite data over Arabian sea, North Atlantic Ocean, Canada, Russia etc. Both the models has given the minimum difference of SWupRF within the selected wavelength of each gas. The overall analysis within the SW-NIR band has been shown an ability to compute the spectral radiative flux over different atmospheric surfaces and can be used to calculate the effects of increases of greenhouse gases. The detailed investigation has been required to add cloud features by comparing the radiance enhancement over different spatial locations with the radiative effect of SW-NIR range. This will definitely reduce the quantification process of detection of cloud scenes and its relationships with concentrations of water vapor and $CO_2$ and will also helpful to extrapolate the consequence of full SW wavelength range in contrast with the climate behavior.

## Acknowledgments

This study was technically supported by Physics & Astronomy department of York University, Canadian Advanced Nano space Experiment 2 Operations Team and Thoth Technologies, Inc. The authors would like to thanks to all of the organizations for the support and their work in operating the spacecraft and also providing all the background information and observed data from space.

## References:

1. Hatzianastassiou, N., Matsoukas, C., Fotiadi, A., Pavlakis, K. G., Drakakis, E., Hatzidimitriou, D., & Vardavas, I. (2005). Global distribution of Earth's surface shortwave radiation budget. Atmospheric Chemistry and Physics, 5(10), 2847-2867.

2. Field, C. B., Barros, V. R., Dokken, D. J., Mach, K. J., Mastrandrea, M. D., Bilir, T. E., ... & Girma, B. (2014). IPCC, 2014: Climate Change 2014: Impacts, Adaptation, and Vulnerability. Part A: Global and Sectoral Aspects. Contribution of Working Group II to the Fifth Assessment Report of the Intergovernmental Panel on Climate Change.

3. Stephens, G. L., Li, J., Wild, M., Clayson, C. A., Loeb, N., Kato, S., & Andrews, T. (2012). An update on Earth's energy balance in light of the latest global observations. Nature Geoscience, 5(10), 691-696.

4. Andrews, T., Forster, P. M., & Gregory, J. M. (2009). A surface energy perspective on climate change. Journal of Climate, 22(10), 2557-2570.

5. Wang, T., Yan, G., Shi, J., Mu, X., Chen, L., Ren, H., ... & Zhao, J. (2014, July). Topographic correction of retrieved surface shortwave radiative fluxes from space under clear-sky conditions. In 2014 IEEE Geoscience and Remote Sensing Symposium (pp. 1813-1816). IEEE.

6. Tang, B., Li, Z. L., & Zhang, R. (2006). A direct method for estimating net surface shortwave radiation from MODIS data. Remote Sensing of Environment, 103(1), 115-126.

7. Inamdar, A. K., & Guillevic, P. C. (2015). Net Surface Shortwave Radiation from GOES Imagery - Product Evaluation Using Ground-Based Measurements from SURFRAD. Remote Sensing, 7(8), 10788-10814.

8. Mitchell, D. L., & Finnegan, W. (2009). Modification of cirrus clouds to reduce global warming. Environmental Research Letters, 4(4), 045102.

9. Mateos, D., Antón, M., Valenzuela, A., Cazorla, A., Olmo, F. J., & Alados-Arboledas, L. (2014). Efficiency of clouds on shortwave radiation using experimental data. Applied Energy, 113, 1216-1219.

10. Furlan, C., De Oliveira, A. P., Soares, J., Codato, G., & Escobedo, J. F. (2012). The role of clouds in improving the regression model for hourly values of diffuse solar radiation. Applied Energy, 92, 240-254.

11. Islam, M. D., Kubo, I., Ohadi, M., & Alili, A. A. (2009). Measurement of solar energy radiation in Abu Dhabi, UAE. Applied Energy, 86(4), 511-515.

12. Barkstrom, B. R. (1984). The earth radiation budget experiment (ERBE). Bulletin of the American Meteorological Society, 65(11), 1170-1185.

13. Barkstrom, B. R., & Smith, G. L. (1986). The earth radiation budget experiment: Science and implementation. Reviews of Geophysics, 24(2), 379-390.

14. Wielicki, B. A., Barkstrom, B. R., Baum, B. A., Charlock, T. P., Green, R. N., Kratz, D. P., ... & Young, D. F. (1998). Clouds and the Earth's Radiant Energy System (CERES): algorithm overview. IEEE Transactions on Geoscience and Remote Sensing, 36(4), 1127-1141.

15. Tarpley, J.D. Estimating incident solar radiation at the surface from geostationary satellite data. J. Appl. Meteor. 1979, 18, 1172–1181.

16. Gautier, C.; Diak, G.R.; Masse, S. An investigation of effects of the spatially averaging satellite brightness measurements on the calculation of insolation. J. Clim. Appl. Meteor. 1984, 23, 1380–1386.

17. Diak, G.R.; Gautier, C. Improvements to a simple physical model for estimating insolation from GOES data. J. Clim. Appl. Meteor. 1983, 22, 505–508.

18. Darnell, W.L.; Staylor, W.F.; Gupta, S.K.; Denn, F.M. Estimation of surface insolation using sun-synchronous satellite data. J. Clim. 1988, 1, 820–835.

19. Pinker, R.T.; Lasszlo, I. Modeling surface solar irradiance for satellite applications on global scale. J. Appl. Meteor. 1992, 31, 194–211.

20. Perez, R.; Ineichen, P.; Moore, K.; Kmiecik, M.; Chain, C.; George, R.; Vignola, F. A new operational satellite-to-irradiance model—Description and validation. Sol. Energy 2002, 73, 307–317.

21. Zheng, T.; Liang, S.; Wang, K.C. Estimation of incident PAR from GOES imagery. J. Appl. Meteor. Clim. 2008, 47, 853–868.

22. Schmetz, J. Towards a surface radiation climatology. Retrieval of downward irradiance from satellites. Atmos. Res. 1989, 23, 287–321.

23. Schmetz, J. Retrieval of surface radiation fluxes from satellite data. Dyn. Atmos. Ocean. 1991, 16, 61–72.

24. McCoy, D. T., Hartmann, D. L., & Grosvenor, D. P. (2014). Observed Southern Ocean Cloud Properties and Shortwave Reflection. Part II: Phase Changes and Low Cloud Feedback*. Journal of Climate, 27(23), 8858-8868.

25. Henderson, D. S., L'Ecuyer, T., Stephens, G., Partain, P., & Sekiguchi, M. (2013). A multisensor perspective on the radiative impacts of clouds and aerosols. Journal of Applied Meteorology and Climatology, 52(4), 853-871.

26. Jiang, J. H., Su, H., Zhai, C., Perun, V. S., Del Genio, A., Nazarenko, L. S., & Gettelman, A. (2012). Evaluation of cloud and water vapor simulations in CMIP5 climate models using NASA "A-Train" satellite observations. Journal of Geophysical Research: Atmospheres, 117(D14).

27. Jagpal, R. K., Quine, B. M., Chesser, H., Abrarov, S. M., & Lee, R. (2010). Calibration and in-orbit performance of the Argus 1000 spectrometer-the Canadian pollution monitor. Journal of Applied Remote Sensing, 4(1), 049501-049501.

28. Siddiqui, R., Jagpal, R., Salem, N. A., & Quine, B. M. (2015). Classification of cloud scenes by Argus spectral data. International Journal of Space Science and Engineering, 3(4), 295-311.

29. Quine, B. M., & Drummond, J. R. (2002). GENSPECT: a line-by-line code with selectable interpolation error tolerance. Journal of Quantitative Spectroscopy and Radiative Transfer, 74(2), 147-165.

30. Chesser, H., Lee, R., Benari, G., Jagpal, R., Lam, K., & Quine, B. (2012). Geolocation of Argus flight data. IEEE Transactions on Geoscience and Remote Sensing, 50(2), 357-361.

31. Sarda, Karan, Stuart Eagleson, Eric Caillibot, Cordell Grant, Daniel Kekez, Freddy Pranajaya, Robert E. Zee "Canadian advanced nanospace experiment 2: Scientific and technological innovation on a three-kilogram satellite." Acta Astronautica 59.1 (2006): 236–245.

32. Argus 1000 IR Spectrometer, Owner's Manual, Thoth Technology Inc. Document Number OG728001, Release 1.03

33. Tsouvaltsidis, C., Benari, G., Al Salem, N.Z., Quine, B. and Lee, R., 2015. ArgusE: Design and Development of a Micro-Spectrometer used for Remote Earth and Atmospheric Observations. In Proceedings of the Advanced Maui Optical and Space Surveillance Technologies Conference, held in Wailea, Maui, Hawaii, September 15-18, 2014, Ed.: S. Ryan, The Maui Economic Development Board, id. 98 (Vol. 1, p. 98).

34. Jagpal, R.K. (2011) Calibration and Validation of Argus 1000 Spectrometer – A Canadian Pollution Monitor, PhD thesis, York University, Canada.

35. Reuter, M., Buchwitz, M., Schneising, O., Heymann, J., Bovensmann, H. and Burrows, J.P., 2010. A method for improved SCIAMACHY $CO_2$ retrieval in the presence of optically thin clouds. Atmospheric Measurement Technique*s*, *3*(1), pp.209-232.